\documentclass{ws-p8-50x6-00}

\begin{document}

\title{Deuteron production and space-momentum correlations at RHIC}
\author{S.Y. Panitkin}
\address{Department of Physics, Kent State University, OH 44242, USA}
\author{B. Monreal$^{\dag}$, N. Xu}
\address{Nuclear Science Division, LBNL, Berkeley, CA 94720, USA}
\author{W.J. Llope}
\address{T.W. Bonner Nuclear Laboratory, Rice University, TX 77005, USA}
\author{R. Mattiello} 
\address{ Niels Bohr Institute, Blegdamsvej 17, University of
Copenhagen, DK-2100 Copenhagen, Denmark}
\author{H. Sorge} 
\address{ Department of Physics, SUNY at Stony Brook, NY 11794, USA} 


\maketitle

\abstracts{We present predictions for the formation of (anti)nuclear
bound states  in Au+Au central collisions at $\sqrt{s}=200A$ GeV.
The coalescence afterburner was applied to the freeze-out phase space
distributions of nucleons provided by the transport model, RQMD
version 2.4.
We study the sensitivity of the deuteron spectra to space-momentum
correlations. It is found that the deuteron transverse momentum
distributions are strongly affected by the nucleon space-momentum
correlations}    

\section{Introduction}
Nuclear clusters have been a useful tool to establish
collective effects throughout the history of heavy ion reactions:
production rates have provided evidence for low temperature phase
transitions,\cite{ALADIN}  the spectral distribution
shows particular sensitivities to collective
flow,\cite{Lisa,E802,E877} transverse 
expansion\cite{UHeinz,Nagle,PRL,Polleri,NA44} and potential
forces.\cite{PRL,Danielewiczs}  
With planned commisioning of the Relativistic Heavy-Ion Collider (RHIC) the
necessity of the predictions of the baryon distribution in
general, and light clusters in particular, is evident.
It should be mentioned that predictions of different transport models
for RHIC energies already offer large differences
in rather basic observables like total particle multiplicities, etc.
 In this paper we present some results of calculations based on the
cascade model RQMD version 2.4\cite{Sorge95} and a coalescence
afterburner.\cite{PRL,Nagle,Sorge1} These results are part of the
effort to formulate a physics program for the STAR collaboration. More
details related to this study and extensive discussion of the
sensitivity of the light nuclei to various properties of the
colliding system at RHIC can be found elsewhere.\cite{rhic_d_1,rhic_d_2} 
\section{Rapidity Distributions}
One of the basic observables in nucleus-nucleus collisions is rapidity
distributions of nucleons. It reflects the energy loss of the
nucleons and as well as bulk properties of the particle production in
a collision. Figure {1} shows predictions, based on RQMD calculations
of rapidity distributions of protons and deuterons as well as
antiprotons and antideuterons for central Au+Au collisions at full
RHIC energy. As has been mentioned earlier, predictions for clusters
were made in a coalescence framework. Vertical dashed lines on 
Figure 1 schematically show the expected acceptance of the STAR TPC.  One
can conclude from the figure that RQMD predicts 1 deuteron to be
emmited into the STAR acceptance for about every 20 central
events. The predicted rate of antideuteron production 
is about 1 per 100 events. With the expected trigger rate of STAR for
Au+Au central collisions ($\sim$1 Hz) these pedicted rates make
deuteron and antideuteron measurements feasible and
a good candidate for a ``year one'' physics.  
\begin{figure}
\centerline{\epsfig{figure=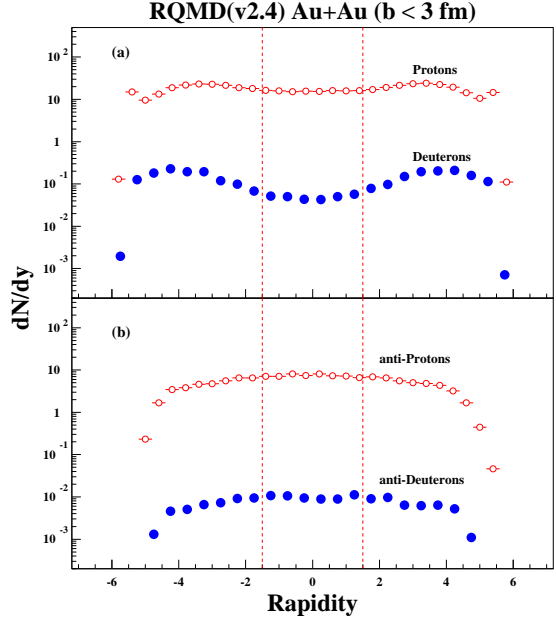,width=8.0cm}}
\caption{ Rapidity distributions of (a) protons (open circles) and
deuterons (filled circles); (b) anti-protons (open circles) and
anti-deuterons (filled circles), from central Au+Au collisions. The
STAR TPC acceptance is indicated by the dashed lines.}
\end{figure}
\section{Transverse Momentum Distributions}
Another important basic observable of the heavy-ion collision is
the transverse momentum distribution of baryons. It is sensitive to 
various physical properties of the collision. 
Transverse momentum distributions reflect the degree of
thermalization reached in the heavy-ion collision as well as effects
of the collective flow.
Fig. 2(a) presents the rapidity dependence of the average transverse
momentum of protons and deuterons for 
normal RQMD events. The average transverse momentum for deuteron is
about a factor of two higher than for protons. For comparison,
rapidity dependence of the mean transverse 
momentum of pions and kaons is shown as dashed and solid lines
respectively. A clear ``particle mass hierarchy'' of the mean $p_T$ is
evident from Fig. 2(a), which is commonly\cite{NA44} attributed to the
presence of the transverse flow.
\begin{figure}
\centerline{\epsfig{figure=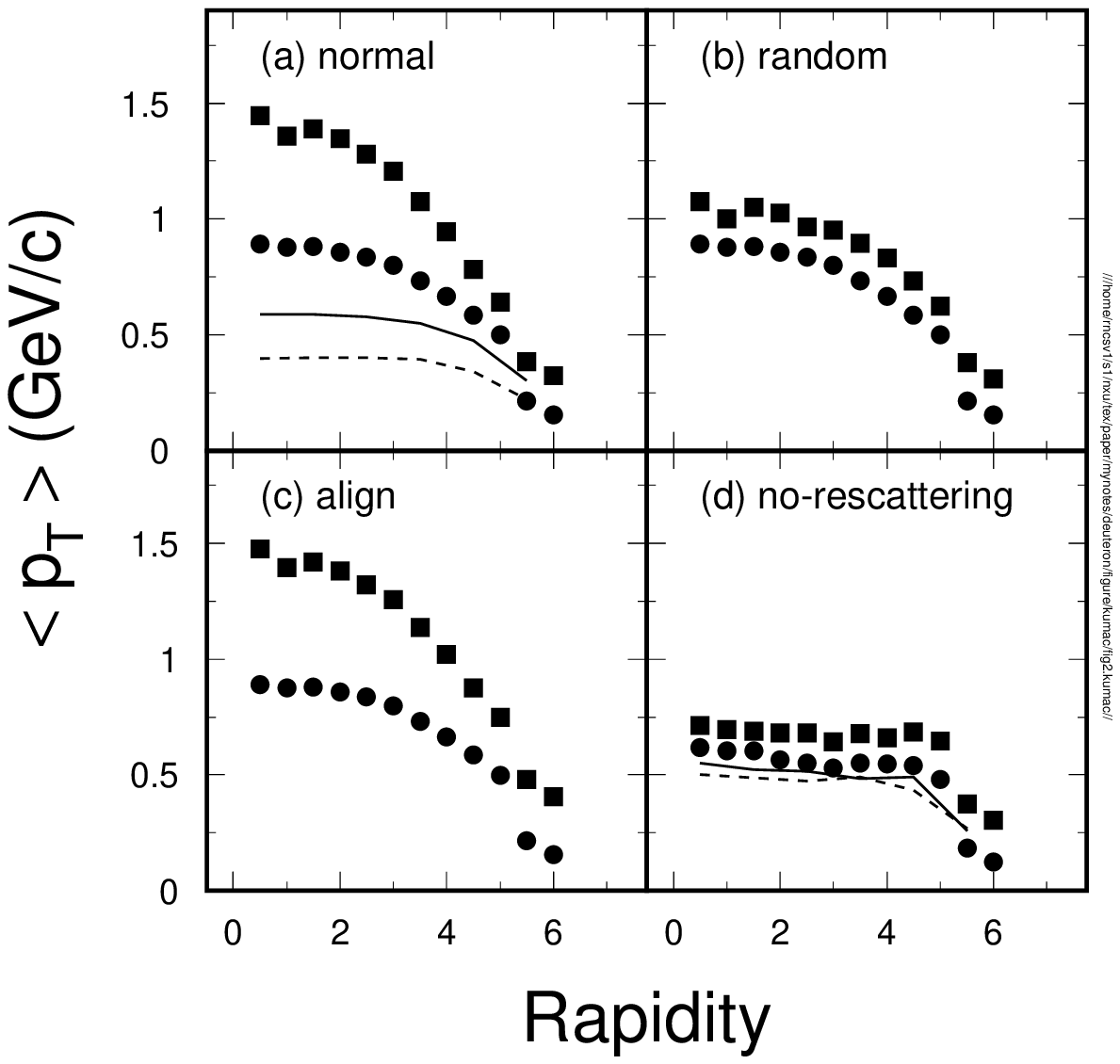,height= 8.cm, width=8.cm}}
\caption{ Deuteron (filled square), nucleon (filled circle), kaon
(solid-line), and pion (dashed-line) mean transverse momentum as a
function of rapidity.}
\end{figure}
The influence of the collective transverse flow component on the mean
transverse momentum can be further demonstrated by comparing this result
with calculations where the freeze-out correlation of positions and
momenta of the nucleons have been deliberately altered.
Panels labeled (b) and (c) in Fig. 2 show the result of such calculations.  
Fig. 2(b) shows the case where the angle between $\vec{p_T}$ and
$\vec{r_T}$ has been randomized. This procedure produces a system
with no collective flow. It can be seen from the figure
that the difference between the average transverse momenta of
deuterons and protons is dramatically reduced.
 Fig. 2(c) shows the so called aligned case, where for each nucleon
 the transverse radius vector $\vec{r_T}$ is aligned with the
 transverse momentum  vector $\vec{p_T}$. This case mimics a
``maximum flow'' scenario. Note that  in the aligned and random cases
only the relative orientation of  $\vec{r_T}$ to $\vec{p_T}$ is
modified: momentum distributions and projections onto either ${r_T}$
or ${p_T}$ are not touched. These figures illustrates high sensitivity
of the deuteron spectra to the momentum position correlations. Figure 2
(d) is the result 
 of a calculation without rescattering among baryons (rescattering
 here means interaction with produced particles). Similar to
 the random case, the results of calculations without baryon
 rescattering (Fig. 2(d)) show a constant, rapidity independent
difference between deuteron and proton transverse momentum of about
150 MeV. This suggests that multiple rescattering among particles in
RQMD leads to collective flow.\cite{rhic_d_2} 

\indent  In summary, using a microscopic transport model RQMD and a
 coalescence afterburner, we have calculated rapidity distributions of
 protons and deuterons as well as their antiparticles for central
 Au+Au collisions at $\sqrt{s} =  200A$ GeV. We studied the   
 sensitivity of the deuteron transverse momentum
 distributions in different  rapidity regions to the effects of the
 transverse collective flow. Should
 new physics occur at RHIC, a modification of the
 space-momentum structure will manifest itself in the deuteron yields
 and transverse momentum distributions. These distributions can be
 measured in the STAR TPC and other RHIC experiments.\\
\indent {$\dag$ B. Monreal is at Lawrence Berkeley National Laboratory through
the Center for Science and Engineering Education.}
\section*{Acknowledgement}
 We are grateful for many enlightening discussions with
 Drs. S. Johnson, D. Keane, S. Pratt, H.G. Ritter, S. Voloshin,
 R. Witt. We  especially thank Dr. J. Nagle for permission to use his
 coalescence code. This research used resources of the National
 Energy Research Scientific Computing Center.  This work has been
 supported by the U.S. Department of Energy under Contract
 No. DE-AC03-76SF00098 and W-7405-ENG-36, DOE grant DE-FG02-89ER40531
 and the Energy Research Undergraduate Laboratory Fellowship and National Science
 Foundation.


\end{document}